\documentclass[pra,twocolumn,showpacs,superscriptaddress,amsmath,amssymb]{revtex4}
\usepackage{mathrsfs}

\usepackage{graphicx}

\usepackage{color}

\begin{document}

\title{An anyon model in a toric honeycomb lattice }

\author{Long Liang}
\affiliation{Institute of Theoretical Physics, Chinese Academy of
Sciences, P.O. Box 2735, Beijing 100190, China}

\affiliation{Department of Physics, East China Normal University,
Shanghai 20024, China }

\author{Yue Yu}
\affiliation{Institute of Theoretical Physics, Chinese Academy of
Sciences, P.O. Box 2735, Beijing 100190, China}

\begin{abstract}
{We study an anyon model in a toric honeycomb lattice. The ground
states and the low-lying excitations coincide with those of Kitaev
toric code model and then the excitations obey mutual semionic
statistics. This model is helpful to understand the toric code of
anyons in a more symmetric way. On the other hand, there is a direct
relation between this  toric honeycomb model and a boundary coupled
Ising chain array in a square lattice via Jordan-Wigner
transformation. We discuss the equivalence between these two models
in the low-lying sector and realize these anyon excitations in a
conventional fermion system.}
\end{abstract}

\date{today}

 \maketitle

\section{Introductions}

 Kitaev toric code model opens a new
direction to study the storage of quantum information and design of
quantum computation in a topological protected way \cite{kitaev}.
The objects carrying the quantum information are so-called anyons,
the quasiparticles obeying exotic statistics in two-dimensional
condensed matter systems \cite{anyon,wil}.

The first realistic anyonic quasiparticle is Laughlin
quasiparticle in $\nu=1/3$ fractional quantum Hall system
\cite{lau,sn1,sn2} and  possibly obey $\theta=\pi/3$ fractional
statistics \cite{aro,anyonex}. Nowadays, the researches for anyons
mainly focus on the following aspects: non-abelian fractional
quantum Hall states \cite{mr,rr,free}; topological insulator
carrying Majorana fermionic edge mode \cite{FK}; and Kitaev
lattice spin models, i.e., the toric code model \cite{kitaev},
Levin-Wen model \cite{lw} and honeycomb-lattice spin model
\cite{kitaev1}.

 The abelian anyons in Kitaev models have currently attracted many research
interests. Experimentally exciting, manipulating and detecting
abelian anyons have been suggested or tried for the toric code
model \cite{han} and for the insulating phase of Kitaev
honeycomb-lattice model \cite{zd}. The explicit presentation of
nonabelian anyons has also been studied \cite{ville}.

The realization of Kitaev toric code and honeycomb lattice models
is not easy because of the unconventional interaction between
spins. In this paper, we propose a toric honeycomb lattice model
which is equivalent to Kitaev toric code but in a more symmetric
way. It was known that the insulator phase of Kitaev honeycomb
lattice model is equivalent to Kitaev toric code model
\cite{kitaev1} according to a perturbation analysis.

Instead of such a complicated way, in this paper, we directly use
the group element of $\mathbb{Z}_2$ gauge group in  Kitaev honeycomb
model to construct our model Hamiltonian. There are two independent
operators to represent the $\mathbb{Z}_2$ group. They play roles of
the stabilizer operators of the toric code. Thus, we find that our
model is equivalent to Kitaev toric code model but the Hamiltonian
is more symmetric. All the eigenstates can be known because the
model is exactly solvable. The low-lying excitations contain a
closed subset in which the excitations obey the mutual semionic
statistics.

Furthermore, we show that this toric honeycomb lattice model can be
mapped to a decoupled Ising chain array in a square lattice under a
special external  field. The latter has been studied before by one
of us with Li \cite{yuli} and the low-lying excitations of that
model also obey mutual semionic statistics. However, the decoupled
Ising chain array has much higher degrees of freedom of the ground
states and low-lying excitations than those of the honeycomb model.
This makes the difficulty to peel out a given set of the anyons
\cite{yuli}. Carefully checking the mapping, one sees that this is
caused by the lost of the couplings between the chains when a
Jordan-Wigner transformation is used in the mapping. The source of
the lost of the coupling arises possibly from the unproper treatment
of the periodic boundary condition in Jordan-Wigner transformation.
An additional boundary Hamiltonian should be added \cite{yaok},
which is a bit complicated. Focussing only on the low-lying sector,
we can introduce vertical Ising ferromagnetic couplings in two given
vertical chains (the 'boundary chains') to approximate the boundary
couplings. After modifying the totally decoupled chain array model
to the boundary coupled chain array model, the degeneracy of the
ground states becomes consistent with the honeycomb model and so are
the low-lyings excitations. Therefore, this boundary coupled Ising
chain array model is equivalent to the toric honeycomb model in the
low-lying sector.

\begin{figure}[htb]
\begin{center}

\includegraphics[width=5cm]{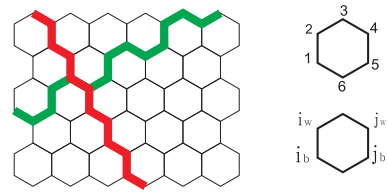}
\end{center}
 \caption{\label{fig:Fig.1} (Color online) Toric honeycomb lattice. The green loop
 (armchair) and the red loop (zig-zag) are two independent
 non-trivial loops.
 }
\end{figure}

\section{ Toric honeycomb model }

 The model we propose is very
simple. Consider a honeycomb lattice in a torus (See Fig.
\ref{fig:Fig.1}). The number of total sites is $2N$ and each site is
occupied by a 1/2-spin. Define two plaquette operators
\begin{eqnarray}
&& W_P=\sigma^y_1\sigma_2^x\sigma^z_3\sigma^y_4\sigma_5^x\sigma^z_6,\nonumber\\
&&\tilde
W_P=\sigma^x_1\sigma_2^y\sigma^z_3\sigma^x_4\sigma_5^y\sigma^z_6,
\end{eqnarray}
where 1, $\cdots$, 5, and 6 are the site indices of a plaquette $P$
as shown in Fig. \ref{fig:Fig.1}. It is easy to directly check
$W^2_P=\tilde W^2_P=1$, i.e, the eigenvalues of them are $\pm1$; and
$[W_P,\tilde W_{P'}]=0$ for all plaquette. The Hamiltonian is
simply defined by sum over all plaquette
\begin{eqnarray}
H=-J_1\sum_P  W_P-J_2\sum_P \tilde W_P,
\end{eqnarray}
where $J_1>0$ and $J_2>0$ and thus $W_P$ and $\tilde W_{P'}$ are
two-independent integrals of motion. This spin model is invariant in
$x\,y\,z$ permutation. This model is exactly soluble: the ground
states are given by the states with $W_P=\tilde W_P=1$ for all
plaquette and the excited states are of at one plaquette having
$W_P=-1$ or $\tilde W_P=-1$, which will be studied later in details.

\section{ Ground states and degeneracy}

 The ground states are of the
form
\begin{eqnarray}
|G\rangle=\prod_{P}(1+ W_P)(1+\tilde W_P)|\phi\rangle,
\end{eqnarray}
where $|\phi\rangle$ is an arbitrary reference state, e.g.,
$|\phi\rangle=|\uparrow\cdots\uparrow\rangle$ with each `$\uparrow$'
denoting the eigenvalue 1 of $\sigma^z_i$. The number of
$|\phi\rangle$ is $2^{2N}$, the dimensions of the whole Hilbert
space $\mathscr{H}$. The above construction of the ground states
means the ground state space $L=\mathscr{H}/G$ where $G$ is the
group generated by all independent $W_P$ and $\tilde W_P$. On the
torus, due to the periodic boundary condition, there are two
constraints $\prod_PW_P=\prod_P\tilde W_P=1$ for the generators,
which means $G\cong(\mathbb{Z}_2)^{2N-2}$. Therefore, according to
the general theory given by  Gottesman \cite{Go}, the degeneracy of
the ground states are $2^{2N-(2N-2)}=4$, which is exactly the same
as that of the Kitaev toric code model \cite{kitaev}.

One can even make a more intuitional proof similar to what Kitaev
did \cite{kitaev}. For a plaquette $P$, one defines $K_{ij}$ to be
$\sigma^{z}_i\sigma_j^{z}$ for $(i,j)=(1,2),(4,5)$,
$\sigma^{x}_i\sigma_j^{x}$ for $(i,j)=(2,3),(5,6)$, and
$\sigma^{y}_i\sigma_j^{y}$ for $(i,j)=(3,4),(6,1)$; $\tilde K_{ij}$
to be $\sigma^{z}_i\sigma_j^{z}$ for $(i,j)=(1,2),(4,5)$,
$\sigma^{y}_i\sigma_j^{y}$ for $(i,j)=(2,3),(5,6)$, and
$\sigma^{x}_i\sigma_j^{x}$ for $(i,j)=(3,4),(6,1)$. For any closed
loop, one can define two different operators
\begin{eqnarray}
C=K_{ij}K_{jk}\cdots K_{si}, ~~\tilde C=\tilde K_{ij}\tilde
K_{jk}\cdots \tilde K_{si}.
\end{eqnarray}
For any trivial loop on the torus, $C$ and $\tilde C$ are identical
to the identity. There are only two unequivalent non-trivial loops
as shown by the green and red in Fig. \ref{fig:Fig.1}. Along the
green loop,
$C_g=\sigma^y_{i}\sigma^y_{j}\sigma_{k}^z\sigma^z_l\sigma^y_{m}\sigma^y_{n}
\sigma_{s}^z\sigma^z_t\cdots$ and $\tilde C_g=\sigma^x_{i}\sigma^x_{j}\sigma_{k}^z\sigma^z_l\sigma^x_{m}\sigma^x_{n}
\sigma_{s}^z\sigma^z_t\cdots$ while along the red loop, they are
$C_r=\sigma^x_{i}\sigma^x_{j}\sigma_{k}^x\sigma^x_l\sigma^x_{m}\sigma^x_{n}
\sigma_{s}^x\sigma^x_t\cdots$ and $\tilde
C_r=\sigma^y_{i}\sigma^y_{j}\sigma_{k}^y\sigma^y_l\sigma^y_{m}\sigma^y_{n}
\sigma_{s}^y\sigma^y_t\cdots$. These four operators generates all
linear operators acting on the ground state space. In this way, we
explicitly prove that the ground states are fourfold degenerate.

\section{ Low-lying excitations}

\subsection{Excitations}

\begin{figure}[htb]
\begin{center}
\includegraphics[width=7cm]{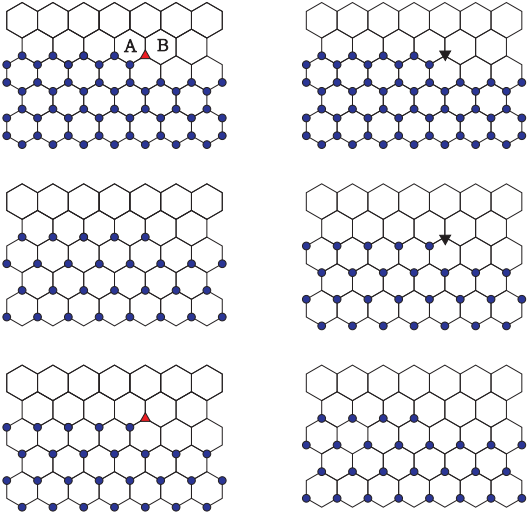}
\end{center}
 \caption{\label{fig:Fig.2} (Color online) The low-lying excitations.
  From left to right and up to down, they are
 $\psi_{i_{b}}$, $\chi_{i_{b}}$, $\sigma^{(1)}_{P}$, $\sigma^{(2)}_{P,P'}$,
  $\sigma^{(3)}_{P,P'}$, and $\sigma_{P'}^{(1)}$. The up-triangles
  (red), down-triangles (black) and blue circles indicate
 acting the operators $\sigma^x$,
  $\sigma^y$ and $\sigma^z$ at these sites on the ground state, respectively.
  }
\end{figure}

Low-lying excitations of this exactly soluble model are given by
\begin{eqnarray}\label{equation1}
&&\sigma^{z}_{i_{b}}|G\rangle,\nonumber\\
&&\psi_{i_{b}}|G\rangle=\sigma^{x}_{i_{b}}\prod_{i'_{s}<i_{b}}
\sigma^{z}_{i'_{s}}|G\rangle,\nonumber\\
&&\chi_{i_{b}}|G\rangle=\sigma^{y}_{i_{b}}\prod_{i'_{s}<i_{b}}
\sigma^{z}_{i'_{s}}|G\rangle,\nonumber\\
&&\sigma^{(1)}_{P}|G\rangle=\sigma^{z}_{i_{b}}\sigma^{z}_{i_{b}-2}
\sigma^{z}_{i_{b}-4}\cdots|G\rangle,\nonumber\\
&&\sigma^{(2)}_{P,P'}|G\rangle=\sigma^{y}_{i_{b}}\sigma^{z}_{i_{b}-1}
\sigma^{z}_{i_{b}-3}\cdots|G\rangle,\nonumber\\
&&\sigma^{(3)}_{P,P'}|G\rangle=\sigma^{x}_{i_{b}}\sigma^{z}_{i_{b}-1}
\sigma^{z}_{i_{b}-3}\cdots|G\rangle,\nonumber\\
&&\sigma^{(1)}_{P'}|G\rangle=\sigma^{z}_{i_{b}-2}\sigma^{z}_{i_{b}-4}
\sigma^{z}_{i_{b}-6}\cdots|G\rangle,
\end{eqnarray}
where the sites $i_b$ are shown in Fig. \ref{fig:Fig.1} and $P$ ($P'$) label the plaquette where the site '1'=$i_b$
($i_b-2$); and the order of the sites is defined as follows:
$i_s>j_t$ if the zig-zag line including $i_s$ is higher than that of
$j_t$ or if $i_s$ is on the right hand of $j_t$ when they are in the
same line. Except the first local excitation, all others are
string-like excitations which are plotted in Fig.\ref{fig:Fig.2}.

 Notice that $\sigma^x$, $\sigma^y$ and $\sigma^z$ anticommute with
each other and therefore, if there is a red triangle ($\sigma^x$) or
black triangle ($\sigma^y$) at site 3 or site 6 in a plaquette $P$,
the eigenvalues of $W_P$ and $\tilde{W}_P$ of this plaquette become
$-1$; if there is a blue circle ($\sigma^z$) at site 1, site 2, site
4 or site 5 in $P$, the eigenvalues of $W_P$ and $\tilde{W}_P$ of
this plaquette also become $-1$; if there is a red triangle at site
1 or site 4, or a black triangle at site 2 or site 5, the eigenvalue
of $W_P$ becomes $-1$; if there is a black triangle at site 1 or
site 4, or a red triangle at site 2 or site 5, the eigenvalue of
$\tilde{W}_P$ becomes $-1$. Taking $\psi_{i_{b}}$ as an example, the
blue circle at site 1 in plaquette A (See Fig.\ref{fig:Fig.2}) makes
the eigenvalues of $W_A$ and $\tilde{W}_A$ become $-1$, while the
red triangle at site 5 makes the eigenvalue of $\tilde{W}_A$ become
$-1$, and finally, the eigenvalue of $W_A$ in plaquette A becomes
$-1$. For plaquette B, the red triangle at site 1 makes the
eigenvalue of $W_B$ become $-1$. It is easy to check that the
eigenvalues of $W$ and $\tilde W$ in other plaquette are not
changed.

Although string-like excitations look like non-local , their
energies are located at one or two sites. The energies of
excitations can be calculated directly, which in turn are $4J_{1}$,
$4J_{2}$, $4J_{1}+4J_{2}$ for $\psi,\chi,\sigma^z$ and
$2J_{1}+2J_{2}$ for $\sigma^{(1,2,3)}$, respectively.

Notice that changing the subscript $i_b$ to $i_{w}$, the excitations
do not change, i.e., we can redefine excitations as follows
\begin{eqnarray}\label{equation2}
&&\sigma^{z}_{i_{w}}|G\rangle,\nonumber\\
&&\psi_{i_{w}}|G\rangle=\sigma^{y}_{i_{w}}\prod_{i'_{s}<i_{w}}
\sigma^{z}_{i'_{s}}|G\rangle,\nonumber\\
&&\chi_{i_{w}}|G\rangle=\sigma^{x}_{i_{w}}\prod_{i'_{s}<i_{w}}
\sigma^{z}_{i'_{s}}|G\rangle,\nonumber\\
&&\sigma^{(1)}_{P}|G\rangle=\sigma^{z}_{i_{w}}\sigma^{z}_{i_{w}-2}
\sigma^{z}_{i_{w}-4}\cdots|G\rangle,\nonumber\\
&&\sigma^{(2)}_{P,P'}|G\rangle=\sigma^{x}_{i_{w}}\sigma^{z}_{i_{w}-1}
\sigma^{z}_{i_{w}-3}\cdots|G\rangle,\nonumber\\
&&\sigma^{(3)}_{P,P'}|G\rangle=\sigma^{y}_{i_{w}}\sigma^{z}_{i_{w}-1}
\sigma^{z}_{i_{w}-3}\cdots|G\rangle,\nonumber\\
&&\sigma^{(1)}_{P'}|G\rangle=\sigma^{z}_{i_{w}-2}\sigma^{z}_{i_{w}-4}
\sigma^{z}_{i_{w}-6}\cdots|G\rangle.
\end{eqnarray}
The only difference is the interchange between $\sigma^{x}$ and
$\sigma^{y}$, and it is easy to check excitations defined in Eqs.
(\ref {equation1}) are the same as those in Eqs. (\ref {equation2}).

\subsection{Fusion Rules}

All excitations are explicitly expressed by Pauli matrices. The
fusion rules of these operators then can be directly calculated (See
TABLE.\ref{table1}).
%\vspace{5mm}

\def\arraystretch{0}

\begin{table}[htb] \caption{\label{table1}Fusion rules of
excitations.}
\begin{tabular}{|c|c|c|c|c|c|c|c|}
\hline &$\psi_{i_b}$
&$\chi_{i_b}$&$\sigma^z_{i_b}$&$\sigma_P^{(1)}~$&
$\sigma_{P,P'}^{(2)}$
&$\sigma_{P,P'}^{(3)}$ &$\sigma_{P'}^{(1)}~$ \\
\hline  $~\psi_{i_b}~~$ &$~I~$ &$i\sigma^z_{i_b}$ &$-i\chi_{i_b}$
&$-i\sigma_{P,P'}^{(2)}$ & $i\sigma_P^{(1)}~$
 &$\sigma_{P'}^{(1)}~$ &$\sigma_{P,P'}^{(3)}$ \\
\hline  $~\chi_{i_b}~~$ &$-i\sigma^z_{i_b}$  &$~I~$ &$i\psi_{i_b}$
&$i\sigma_{P,P'}^{(3)}$ &$\sigma_{P'}^{(1)}~$
  & $-i\sigma_P^{(1)}~$
 &$\sigma_{P,P'}^{(2)}$ \\
\hline  $~\sigma_{i_b}^z~~$ &$i\chi_{i_b}$ &$-i\psi_{i_b}$ &$~I~$
&$\sigma_{P'}^{(1)}$  & $-i\sigma_{P,P'}^{(3)}$
 &$i\sigma_{P,P'}^{(2)}$ &$\sigma_{P}^{(1)}$~ \\
\hline $\sigma_{P}^{(1)}~$ &$i\sigma_{P,P'}^{(2)}$
&$-i\sigma_{P,P'}^{(3)}$ &$\sigma_{P'}^{(1)}$ &$~I~~$
 & $-i\psi_{i_b}~~$
 &$i\chi_{i_b}~~$ &$\sigma_{i_b}^z~~$  \\
\hline $~\sigma^{(2)}_{P,P'}$ &$-i\sigma_{P}^{(1)}$
&$\sigma_{P'}^{(1)}$ &$i\sigma_{P,P'}^{(3)}$ &i$\psi_{i_b}$
 & $~I~~~$
 &$-i\sigma_{i_b}^z~~$ &$\chi_{i_b}~~$  \\
\hline $~\sigma^{(3)}_{P,P'}$ &$\sigma_{P'}^{(1)}$
&$i\sigma_{P}^{(1)}$ &$-i\sigma_{P,P'}^{(2)}$ &$-i\chi_{i_b}$
 & $i\sigma_{i_b}^z$
 &$~I~~~$ &$\psi_{i_b}~~$  \\
\hline $~\sigma^{(1)}_{P'}$
&$\sigma_{P,P'}^{(3)}$&$\sigma_{P,P'}^{(2)}$ &$\sigma_{P}^{(1)}$
&$\sigma_{i_b}^z$
 & $\chi_{i_b}$
 &$\psi_{i_b}$ &$~~I~~$ \\
\hline
\end{tabular}
\end{table}

%\vspace{5mm}
The fusion rules of the closed subset $\{I, \psi_{i_b},
\sigma_P^{(1)},\sigma^{(2)}_{P,P'}\}$ are exactly the same as the fusion
rules in Kitaev toric code model if we identify $\psi,\sigma^{(1)}$
and $\sigma^{(2)}$ as $\varepsilon, e$ and $m$\cite{kitaev,kitaev1}:
\begin{eqnarray}
&&\psi^2=(\sigma^{(1)})^2=(\sigma^{(2)})^2=1,\nonumber\\
&&\sigma^{(1)}\sigma^{(2)}=\psi,~~\psi\sigma^{(1)}=\sigma^{(2)},~~
\sigma^{(2)}\psi=\sigma^{(1)}.
\end{eqnarray}

Here `= ' may be up to a sign and/or `$i$'.
 Similarly, the subset  $\{I, \chi_{i_b}, \sigma_P^{(1)},
\sigma^{(3)}_{P,P'}\}$ also obeys the same fusion rules.

Because of the periodic boundary condition for a torus, the number of
string-like excitations must be even.

\section{ Braiding rules and anyons}

 Enlightened by this equivalence
in the fusion rules, we check the braiding matrix. At first we find
the statistics of the same kind excitations. It is easy to get
$[\sigma^{(1)}_{P},\sigma^{(1)}_{P'}]=0$,
$[\sigma^{(2)}_{P_1,P'_1},\sigma^{(2)}_{P_2,P'_2}]=0$, and
$\{\psi_{i_b},\psi_{j_b}\}=2\delta_{ij}$, i.e. $\sigma^{(1)}_{P}$
and $\sigma^{(2)}_{P,P'}$ obey Bose statistics, while $\psi_{i_{b}}$
obey Fermi statistics.
 Since $\sigma^{(1)}\sigma^{(2)}=\psi$, the interchange between two
 $\psi$, $R_{\psi\psi}$, can be considered as product of
 $R_{\sigma^{(1)}\sigma^{(1)}}$, $R_{\sigma^{(2)}\sigma^{(2)}}$,
 $R_{\sigma^{(1)}\sigma^{(2)}}$ and $R_{\sigma^{(2)}\sigma^{(1)}}$,
 i.e.,\cite{note}
\begin{eqnarray}
R_{\psi\psi}=R_{\sigma^{(1)}\sigma^{(1)}}R_{\sigma^{(2)}\sigma^{(2)}}
R_{\sigma^{(1)}\sigma^{(2)}}R_{\sigma^{(2)}\sigma^{(1)}}.
\end{eqnarray}
Clearly, $R_{\psi\psi}=-1$, $R_{\sigma^{(1)}\sigma^{(1)}}=1$ and
$R_{\sigma^{(2)}\sigma^{(2)}}=1$, so
$R_{\sigma^{(1)}\sigma^{(2)}}R_{\sigma^{(2)}\sigma^{(1)}}=-1$. This
means when $\sigma^{(1)}$ circles around $\sigma^{(2)}$ or vice
versa, a minus sign is acquired. In other words, they obey mutual
semionic statistics. Braiding $\psi_{i_{b}}$ with $\sigma^{(1)}$ or
$\sigma^{(2)}$ also gives $-1$. Similarly, statistics among
$\chi_{i_b}$, $\sigma^{(1)}$ and $\sigma^{(3)}$ is also semionic.

\section{Mapping to a square lattice}

\subsection{Jordan-Wigner Transformation}

This model can be mapped to a square lattice via a Jordan-Wigner
transformation. For simplicity, we first restrict to the open
boundary condition. We have defined $\psi_{i_{b}}$, $\chi_{i_{b}}$,
$\psi_{i_{w}}$ and $\chi_{i_{w}}$ before. It is easy to check they
are Hermitian and $\{\psi_{i_{b}},\psi_{j_{b}}\}=2\delta_{ij}$,
$\{\psi_{i_{w}},\psi_{j_{w}}\}=2\delta_{ij}$ and
$\{\psi_{i_{b}},\psi_{j_{w}}\}=0$. These relations are valid for
$\chi$ as well. Furthermore, $\psi$ and $\chi$ are anticommutative. These
operators can be thought as Majorana fermions, and can be combined
into 'complex' fermions
\begin{eqnarray}
c^{\dag}_{\uparrow,i}=\frac{1}{2}(\psi_{i_{w}}-i\psi_{i_{b}}),~~
c_{\uparrow,i}=\frac{1}{2}(\psi_{i_{w}}+i\psi_{i_{b}}),\nonumber\\
c^{\dag}_{\downarrow,i}=\frac{1}{2}(\chi_{i_{w}}-i\chi_{i_{b}}),~~
c_{\downarrow,i}=\frac{1}{2}(\chi_{i_{w}}+i\chi_{i_{b}}).
\end{eqnarray}
Expressing Pauli matrices by $\psi$ and $\chi$, we get
\begin{eqnarray}
&&\sigma^{z}_{i_{w}}=-i\chi_{i_{w}}\psi_{i_{w}},~~
\sigma^{z}_{i_{b}}=-i\psi_{i_{b}}\chi_{i_{b}},\nonumber\\
&&\sigma^{y}_{i_{w}}=\psi_{i_{w}}\prod_{i'_{s}<i_{w}}\sigma^{z}_{i'_{s}},~~
\sigma^{y}_{i_{b}}=\chi_{i_{b}}\prod_{i'_{s}<i_{b}}\sigma^{z}_{i'_{s}},\nonumber\\
&&\sigma^{x}_{i_{w}}=\chi_{i_{w}}\prod_{i'_{s}<i_{w}}\sigma^{z}_{i'_{s}},~~
\sigma^{x}_{i_{b}}=\psi_{i_{b}}\prod_{i'_{s}<i_{b}}\sigma^{z}_{i'_{s}}.
\end{eqnarray}
Thus, $W_{P}$ and $\tilde{W_{P}}$ can be expressed by $\psi$ and
$\chi$ as
\begin{eqnarray}
&&W_{P}=i\psi_{i_{w}}\psi_{i_{b}}i\psi_{j_{w}}\psi_{j_{b}},
\nonumber\\
&&\tilde{W_{P}}=i\chi_{i_{b}}\chi_{i_{w}}i\chi_{j_{b}}\chi_{j_{w}}.
\end{eqnarray}
Using the complex fermions, one has
\begin{eqnarray}
&&W_{P}=(2n_{\uparrow,i}-1)(2n_{\uparrow,j}-1),
\nonumber\\
&&\tilde{W_{P}}=(2n_{\downarrow,i}-1)(2n_{\downarrow,j}-1),
\end{eqnarray}
where $n_{s,i}=c^{\dag}_{s,i}c_{s,i}$ are the fermion number
operators. The Hamiltion can be rewritten as
\begin{eqnarray}
H=-J_1\sum_{\langle
ij\rangle_{hd}}(2n_{\uparrow,i}-1)(2n_{\uparrow,j}-1)\nonumber\\
-J_2\sum_{\langle
ij\rangle_{hd}}(2n_{\downarrow,i}-1)(2n_{\downarrow,j}-1).\label{jwh}
\end{eqnarray}

\begin{figure}[htb]
\begin{center}

\includegraphics[width=7cm]{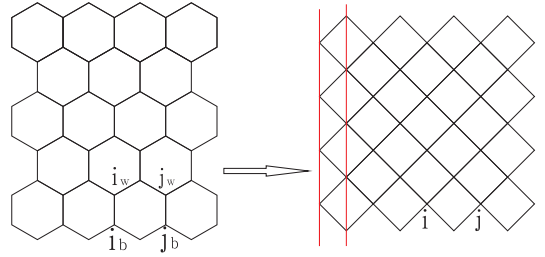}
\end{center}
 \caption{\label{fig:Fig.3}(Color online) Map from
  the honeycomb lattice to the square lattice. Ferromagnetic couplings are added to
  the two red lines.
 }
\end{figure}

The symbol $\langle ij\rangle_{hd}$ means the sum is over nearest
neighbors along the horizontal diagonals of squares (See
Fig.\ref{fig:Fig.3}). In this way, we transfer the honeycomb lattice
model to square lattice model composed of series of decoupled Ising
chains as shown in Fig.\ref{fig:Fig.3}.

\subsection{Ground states and Excitations}

This square lattice model has been studied in our previous work
\cite{yuli}. The ground states of this Hamiltonian are $2^M$-fold
degenerate, i.e., every individual chain is ferromagnetic, i.e., for
a set of spins $\{s_1,\cdots,s_a,\cdots,s_M\}$,
\begin{eqnarray}
|G_{\{s\}}\rangle=\prod_{a=1}^M|G_{s_a}\rangle=\prod_{a,i_a}c^\dag_{s_a,i_a}|0\rangle,
\end{eqnarray}
where $M$ is the number of chains, $i_a$ is the site index in the
$a$-th chain. The low-lying excitations above a given ground state
$|G_{\{s\}}\rangle$, for a given site $i_a$, reads
\begin{eqnarray}
{\cal
H}_{i_a}|G_{\{s\}}\rangle&=&(c^\dag_{s_a,i_a}+c_{s_a,i_a})|G_{\{s\}}\rangle=c_{s_a,i_a}|G_{\{s\}}\rangle,\nonumber\\
{\cal D}_{i_a}|G_{\{s\}}\rangle&=&(c^\dag_{\bar s_a,i_a}+c_{\bar
s_a,i_a})|G_{\{s\}}\rangle
=c^\dag_{\bar s_a,i_a}|G_{\{s\}}\rangle,\nonumber\\
{\cal F}_{i_a}|G_{\{s\}}\rangle&=&i{\cal H}_{i_a}{\cal
D}_{i_a}|G_{\{s\}}\rangle=ic_{s_a,i_a}c^\dag_{\bar
s_a,i_a}|G_{\{s\}}\rangle,\nonumber\\
{\cal W}_P|G_{\{s\}}\rangle&=&\prod_{i'_b\leq i_a} {\cal
F}_{i'_b}|G_{\{s\}}\rangle=\prod_{i'_b\leq i_a}
ic_{s_b,i'_b}c^\dag_{\bar s_b,i'_b}|G_{\{s\}}\rangle,\nonumber\\
{\cal W}_{P,P'}^h|G_{\{s\}}\rangle&=&\prod_{i'_b<i_a} {\cal
F}_{i'_b}{\cal
H}_{i_a}|G_{\{s\}}\rangle\nonumber\\&=&\prod_{i'_b<i_a}ic_{s_b,i'_b}c^\dag_{\bar
s_b,i'_b}c_{
s_a,i_a}|G_{\{s\}}\rangle,\label{exc}\\
 {\cal W}_{P,P'}^d|G_{\{s\}}\rangle&=&\prod_{i'_b<i_a} {\cal
F}_{i'_b}{\cal
D}_{i_a}|G_{\{s\}}\rangle\nonumber\\&=&\prod_{i'_b<i_a}ic_{s_b,i'_b}c^\dag_{\bar
s_b,i'_b}c^\dag_{\bar s_a,i_a}|G_{\{s\}}\rangle,\nonumber
\end{eqnarray}
where $P$ and $P'$ denote two plaquette on the right and left of
$i_a$, respectively. $\bar s=\downarrow(\uparrow)$ if
$s=\uparrow(\downarrow)$. The order of sites is defined by
$i'_b<i_a$ if $i'_b$ is on the left hand of $i_a$ for $b=a$ or
$i'_b$ is in a chain lower than the chain with $i_a$. ${\cal H,D,F}$
create a hole, a double occupant, and a spin-flip. ${\cal W}$,
${\cal W}^d$ and ${\cal W}^h$ create a half-infinite string of
spin-flips, a spin-flip string with a double occupant and a
spin-flip string with a hole, respectively, since the spins of
fermions at sites
 $i'_b<i_a$ are flipped from their ground state configuration while those
at $j_c>i_a$ keep in their ground state configuration. These
excitations are equivalent to those on honeycomb lattices with
\[
\{{\cal H,D,F},{\cal W},{\cal W}^{h},{\cal
W}^d\}\leftrightarrow\{\psi,\chi,\sigma^z,\sigma^{(1)},\sigma^{(2)},\sigma^{(3)}\}.\]

\subsection{Equivalence in Low-lying Sector}

As we have seen, the degeneracy of the ground states of the
decoupled Ising chains is much higher than the toric honeycomb
model. This means two models are not equivalent. This is
caused by unproper treatment of the periodic boundary condition of
the toric lattice in the Jordan-Wigner transformation. Due to the
periodic boundary condition, an additional boundary Hamiltonian
$H_b$ should be added to Eq. (\ref{jwh}) \cite{yaok}. The exact form
of $H_b$ is complicated. Instead, we can use an approximation to
deal with the periodic boundary condition so that the low-lying
excitation sector is correct. We introduce Ising ferromagnetic
boundary couplings to the system, i.e., for the two vertical red
lines in Fig.\ref{fig:Fig.3}, we approximate the boundary effect
from the Jordan-Wigner transformation through a boundary coupling
Hamiltonian
\begin{eqnarray}
H_b=-g_1\sum_{\langle ij\rangle_1;s}n_{is}n_{js}-g_2\sum_{\langle
ij\rangle_2;s}n_{is}n_{js}~,
\end{eqnarray}
where $1$ and $2$ label two vertical red lines in
Fig.\ref{fig:Fig.3} and $\langle ij\rangle_{1,2}$ mean the sum is
over corresponding vertical lines. We assume $g_{1,2}$ is much
larger than $J_{1,2}$, the energy scale of the low-lying
excitations, so that the opposite nearest neighbor spins along a
given boundary line do not create additional low-lying excitations.
Clearly, this kind of couplings make all odd/even chains have the
same states in a given ground state. Therefore, the degeneracy of
the ground states is the same as two decoupled Ising chains, i.e.,
4. And couplings will not change the ground state. Due to the
periodic boundary condition, low-lying excitations will appear in
pairs. Thus, as long as excitations do not cross the vertical
boundary lines, excitations are low-lying . The statistics of
excitations in square lattice are the same as those in honeycomb
lattice, see \cite{yuli}.

\section{Conclusions}

We have constructed a toric honeycomb model and proved that it is
equivalent to Kitaev toric code model in square lattice. This
honeycomb model is helpful to understand the toric code of anyons in
a more symmetric way. It was also shown that this model can be
mapped into an Ising chain array with boundary couplings, which
proposed a possible realization of the anyon models in a
conventional interacting fermion model.

We thank Yi Li for useful discussions and the contribution in the
early stage of this work. This work was supported by National
Natural Science Foundation of China, the national program for basic
research of MOST of China, the Key Lab of Frontiers in Theoretical
Physics of CAS and a fund from CAS.


\begin{references}
\bibitem{note} To have a graphic understanding for this relation, see Fig. 4 in
\cite{yuli}.
\end{references}
\end{document}